\documentclass[preprint2]{aastex}  

\def\msun{$M_{\odot}$}

\def\xte{{\it RXTE~}} 

\slugcomment{Submission Rev. 2B: 2006 May 18}

\begin{document}  
\title{The Rates of Type I X-ray Bursts from Transients Observed with RXTE:
Evidence for Black Hole Event Horizons}


\author{Ronald A. Remillard and Dacheng Lin} 
\affil{MIT Kavli Institute for Astrophysics and Space Research, MIT, 
Cambridge, MA 02139-4307 } 
\email{rr@space.mit.edu; lindc@space.mit.edu}  

\and  

\author{Randall L.\ Cooper and Ramesh Narayan}
\affil{Harvard-Smithsonian Center for Astrophysics, 60 Garden
Street, Cambridge, MA 02138}
\email{rcooper@cfa.harvard.edu, rnarayan@cfa.harvard.edu}
 
\begin{abstract}  

We measure the rates of type I X-ray bursts from a likely flux-limited
sample of 37 non-pulsing Galactic transients observed with RXTE during
1996-2004. These sources are well-categorized in the literature as
either neutron-star systems or black hole candidates.  Our goals are
to test the burst model for neutron stars and to investigate whether
black holes have event horizons. Target selection is one of several
differences between the present study and the investigation of the
event horizon question by Tournear et al. (2003).  Burst rates are
measured as a function of bolometric luminosity, and the results are
compared with augmented versions of the burst model developed by
Narayan \& Heyl (2002; 2003). For a given mass, we consider a range in
both the radius and the temperature at the boundary below the
accretion layer. We find 135 spectrally-confirmed type I bursts in 3.7
Ms of PCA exposures for the neutron-star group (13 sources), and the
burst rate function is generally consistent with the model
predictions. However, for the black hole groups (18 sources), there
are no confirmed type I bursts in 6.5 Ms of exposure, and the upper
limits in the burst function are inconsistent with the model
predictions for heavy compact objects with a solid surface.  The
consistency probability is $\sim 2 \times 10^{-7}$ for dynamical
black-hole binaries, falling to $ 3 \times 10^{-13}$ with the added
exposures of black-hole candidates.  Furthermore, there are systematic
spectral differences between the neutron-star and black-hole groups,
supporting the presumption that physical differences underly the
classifications in our sample. These results provide indirect evidence
that black holes do have event horizons.

\end{abstract}  


\section{Introduction}  

Do black holes really have event horizons as described by general
relativity?  This question has been addressed using several different
methods that judiciously compare the properties of black holes in
accreting binary systems versus accreting neutron stars.  Such
investigations utilize the logic that the hard surface of a neutron
star imposes a final boundary condition on accretion processes, and
that this condition causes observational signatures that may be absent
for black holes with event horizons (see review by Narayan
2003). \nocite{nar03r} The inherent difficulty in gaining inferences
for a negative condition, i.e. that black holes have no surface,
requires that this hypothesis be tested with an accretion process that
can be tied to a detailed and reliable physical model. The confidence
level for the inference of the event horizon can be increased by
investigating this question with as many independent methods as
possible.

Investigations of event horizons may capitalize on the fact that there
are two classes of compact objects, distinguished by mass (Charles \&
Coe 2006; McClintock \& Remillard 2006; Thorsett \& Chakrabarty 1999).
\nocite{cha06, mcc06, tho99} Dynamical studies of the companion
stars in some X-ray binary systems reveal compact objects with masses
(4--14 \msun) that are well above the theoretical upper limit for
neutron stars \citep{kal96}. These are designated as ``dynamical black
hole'' systems.  We adopt this common convention and refer to these
compact objects, henceforth, as black holes (``BH''), although for the
purposes of this paper we consider them to be heavy compact objects
that {\it may} have collapsed within an event horizon.  The other
group consists of neutron stars (``NS''). In practice, compact objects
are routinely identified as NS systems if they display either coherent
pulsations or type I X-ray bursts (see below). Dynamical measurements
of NS binaries are led by the double-pulsar systems detected in the
radio band, supplemented by a handful of X-ray pulsars and bursters
\citep{cha06, tho99}. The results are all consistent with a mass near
1.4 \msun, except for two X-ray pulsars which have an estimated mass
near 2.0 \msun.  Some non-pulsing X-ray sources exhibit the brightness
and variability characteristics of accreting compact objects, but
their type is unknown, with no record of bursts nor measurements of
binary motions, usually because of optically extincted or very faint
companion stars. An important subgroup of such cases are known to
exhibit the X-ray spectral states of black-hole binaries (BHBs), and
they are labeled as black-hole candidates (BHCs), with a confidence
level based on the extent of the observation archive and the clarity
of BH states recorded \citep{mcc06}.

A few different techniques have been used to gain evidence for BH
event horizons. The first method focused on the quiescent state of
X-ray transients, when there is a very low rate of mass transfer from
the donor star to the accretion disk surrounding the compact object.
Quiescent BHs were found to exhibit a systematically lower
X-ray luminosity, compared to quiescent NS transients \citep{nar97,
men99, gar01, mcc04}. An exception to this trend has been found in the
observation of a quiescent NS system with very low luminosity
\citep{jon06}. The inefficient, non-thermal X-ray spectrum for
quiescent BHs was interpreted by a model \citep{nar95} for an
advection-dominated accretion flow (ADAF). This model explains the
lack of a soft thermal component for quiescent BHs
\citep{mcc04}, while the impact of the flow with a hard surface
explains the higher luminosity and thermal X-ray spectrum observed for
quiescent NS systems.  This topic remains under investigation with
evolution in the advection model \citep{nar00, qua00, igu03, nar05}
and controversies about alternative accretion mechanisms such as a
BH jet \citep{fen04}.

More recently, searches for evidence of event horizons have been
initiated for the active states of accretion in X-ray binary systems.
Narayan \& Heyl (2002; 2003) \nocite{nar02,nar03} showed that type I
X-ray bursts can be used to probe the accumulation of accreted mass
onto the surface of a compact object, and they found ways to predict
how the burst rate (versus mass accretion rate) would vary as a
function of the mass of the compact object.  X-ray bursts are detected
as bright and distinctive flares with 10-100 s duration in the X-ray
band. They represent thermonuclear explosions on the surface of a
NS \citep{str06}.  For a wide range of parameters, type I
bursts are an inevitable consequence of the accumulation of matter
that is rich in H and He onto the ultra-dense and hot surface of a
NS, or any hypothesized compact object with a solid surface
and comparable scales of mass and radius.

A third variant of the search for BH event horizons focuses on the
energy spectra of X-ray binaries in full outburst \citep{don03}. When
the patterns of spectral evolution are examined, a distinct type of
soft X-ray spectrum is occasionally observed, but only for BHBs and
BHCs. This spectrum is understood as thermal emission from the inner
accretion disk.  It is argued that NSs with a high accretion rate
cannot exhibit such a simple, low-temperature spectrum because of the
energetics in a boundary layer that forms where accreting matter
impacts the surface. The current level of confidence for the
boundary-layer model may not be very high, given that so many details
of accretion physics are complex and poorly understood.  However,
there do appear to be systematic differences in the X-ray spectra of
accreting NSs vs. BHBs at high luminosity, and these differences are
surely worth investigating in the context of the reality of BH event
horizons.

In this paper, we offer new X-ray data analyses that investigate the
destination of accreted matter by measuring the rates of type I X-ray
bursts (hereafter ``bursts'') in X-ray binary systems.  We compare the
results to augmented versions of the burst model developed by Narayan
\& Heyl (2002; 2003).  A statistical study of burst rates in NS
systems and BHBs was conducted by Tournear et al.~\nocite{tou03}
(2003; ``T03'').  They argue that the mean burst rate for BHBs and
BHCs is indeed far below the rate expected if BHs are heavy compact
objects that do have a hard surface.  In the present paper, we
re-address this question and offer several differences with respect to
the investigation of T03.  1.) Sources are selected from a
flux-limited list of X-ray transients derived from 9 years of
operation by the All-Sky Monitor (ASM) on the {\it Rossi} X-ray Timing
Explorer (\xte). The overlap with T03 for BHBs and BHCs is, in fact,
rather small.  2.) We conduct computer searches for bursts, rather
than ``visual inspection'', in order to carefully screen all burst
candidates and to measure statistical distributions in parameters such
as the detection threshold and the durations of all confirmed and
rejected burst candidates. 3.) We measure the burst rates (and upper
limits) as a function of the unabsorbed bolometric luminosity, scaled
to the Eddington luminosity, in order to most effectively utilize
burst models for both NSs and heavy compact objects.

Physical models for type I X-ray bursts have a rich and complex
history (Woosley \& Taam 1976; Fujimoto, Hanawa, \& Miyaji 1981;
Paczynski 1983; Fushiki \& Lamb 1987; Taam, Woosley, \& Lamb 1996;
Cumming \& Bildsten 2000; Narayan \& Heyl 2002; 2003; Woosley et
al. 2004). \nocite{woo76,fuji81,pac83,fush87,taam96,cum00,nar02,nar03,woo04}
Current theory predicts three different burst regimes that are
delimited by the mass accretion rate.  Abundance changes due to the
stable nuclear burning of H and/or He determine the properties of the
bursts in each regime.  There are some successes and many problems in
applying burst models to explain the distributions in the burst
timescale and amplitude in each regime \citep{kuul02,cor03,str06}.
Nevertheless, there are continuing improvements in the manner in which
stable nuclear burning is handled \citep{nar03,woo04}.  This study requires
a burst model that can reasonably predict the burst rate function for
NS systems, and our sample group of NS transients provides critical
feedback for this purpose.

The three goals of this study are: to determine whether the burst
rates for non-pulsing NS systems match the predictions of the burst
model for NSs, to search for X-ray bursts in BH systems, and to
determine whether the burst rates or upper limits for BHBs and BHCs
are compatible with predictions of the burst model for heavy compact
objects with a solid surface.  In \S 2, we describe our selections of
X-ray targets and \xte pointed observations.  The methods to identify
burst candidates from X-ray timing analysis and to confirm bursts with
X-ray spectral analysis are explained in \S 3.  Details regarding the
burst model are given in \S 4.  The burst rates and comparisons with
the model predictions are given in \S 5, followed by discussions and
conclusions.

\section{Observations}

\subsection {ASM Sample of X-ray Transients, 1996-2004 }

Comparative studies of the BH and NS classes of X-ray binary systems
must contend with the fact that we do not have a large sample of
sources for which there are mass measurements for the compact
objects. This hampers the ability to design statistical studies that
simply define groups on the basis of mass.  The primary
limitation is the small number of NS classifications with dynamical
mass measurements.  One can work around this problem by classifying
sources according to the current paradigms and investigating whether
their behavior is self-consistent with their classification. We adopt
this strategy to investigate accreting compact objects in terms of
X-ray burst production. The selection of some kind of complete
population of X-ray binary systems is then very useful, since this can
help to limit selection biases that may intrude upon the effort to
fairly represent each compact-object class.

We use the archive of the \xte ASM \\
\citep{lev96} over the time
interval 1996 January to 2005 January to define a sample of X-ray
transients whose outbursts are associated with the disk instability
mechanism.  In these sources, matter from the companion star slowly
fills a quiescent accretion disk until a critical density is reached,
and mass then flows to the compact object during an interval of
roughly 1--30 months \citep{che97}.  X-ray transients are selected
because they systematically sweep through a large range in the mass
accretion rate, which is one of the most important variables that
affects the burst rate for a compact object of a given mass.

All of the sources in the ASM archive are considered, except for those
with extragalactic classifications. We use the sum band light curves
(2--12 keV), binned in 1-day time intervals, and we determine three
quantities: the maximum weighted-average flux during any 7-day
interval ($f_{max}$), the highest number of consecutive 7-day
intervals when the source remains detected above a $5 \sigma$
threshold ($N_{det}$), and the maximum number of consecutive 7-day
intervals when the source is below the detection threshold
($N_{off}$). We then apply an ad hoc definition of an X-ray transient
as a source with $3 \le N_{det} \le 150$ and $N_{off} \ge 30$.  The
selection criteria are designed to avoid both fast X-ray transients
(e.g. XTE~J0421+560 and XTE~J1901+031) and quasi-persistent sources
(e.g. KS~1731-260 and XTE~J1716-389). To select a roughly complete
sample, while avoiding all of the systematic complications near the
ASM detection threshold, we further choose a sample with $f_{max} >$
3.0 ASM c/s, i.e., transients with at least one week above 40 mCrab or
$1.0 \times 10^{-9}$ erg cm$^{-2}$ s$^{-1}$ at 2--10 keV. This process
yields 50 X-ray sources. The completeness threshold is estimated from
two considerations.  First, the ASM weekly sky maps have been used to
discover transients (e.g. XTE~J1837+037 and XTE~J0921-319) down to 25
mCrab \citep{gal02}. Second, the PCA scans of the Galactic Center
region provide monitoring functions down to a few mCrab, albeit with
only weekly or bi-weekly sampling.  The PCA scans decrease the
probability that an X-ray transient may go unnoticed in the celestial
area where the ASM detection threshold is highest, due to the large
number of bright sources.

The list of 50 classical X-ray transients contains 9 BHBs and 10 BHCs
\\ \citep{mcc06, oro04, cas04}, 13 X-ray pulsars, 14 known bursters,
and 4 sources that remain unclassified.  The X-ray pulsars (10
classical ones and 3 X-ray msec pulsars) are excluded from our
burst-rate analyses, and the remaining 37 sources are listed in
Table~\ref{tab:xobs}.  The classical X-ray pulsars are excluded
because their strong magnetic field focuses the accretion geometry
onto the magnetic polar caps, where steady nuclear burning suppresses
bursts.  The msec X-ray pulsars are known sources of X-ray
bursts \citep{str03}. However, they may differ from common bursters
because of magnetic field effects and/or the chemical composition of
accreted matter, since the companions are extremely evolved stars or
stripped cores.

Our list of 37 X-ray transients consists of four groups: BHBs, BHCs,
NS-bursters, and ``type unknown''.  Table~\ref{tab:xobs} is organized
in terms of these groups. We note that 5 of the 14 systems in the NS
group do not produce bursts in this investigation (col. 10 of Table
~\ref{tab:xobs}).  However, NS classifications can be made with
contemporaneous instruments, e.g. {\it Beppo}-SAX, or
with other instruments during previous outbursts. 
In particular, X-ray bursts have been recorded from X1711-339
\citep{cor02}, SAX~J1747.0-2853 \citep{wer04}, X1744-361
\citep{eme01}, X1803-245 \\
\citep{mul98, wij99}, and SAX~J1810.8-2609
\citep{nat00}.

Models for X-ray bursts provide predictions of burst rates as a
function of the mass accretion rate, the compact object mass, and a
set of trial assumptions which are discussed in \S4. Predictions can
be tied to observations if it is assumed that the accretion rate
scales with the bolometric luminosity, which can be calculated from
spectral parameters when there are estimates for the distance and the
binary inclination angle (used for the thermal luminosity from the
accretion disk).  Table~\ref{tab:xobs} lists the values for distance
and inclination angle adopted for each source.  The BH masses and
distances are taken from McClintock \& Remillard (2006), while we use
an average mass for the BHC and NS groups.  Distance estimates for the
NS group are primarily derived from published analyses of
radius-expansion bursts \citep{str06}. We assume a distance of 8.5 kpc
for sources with no published distance estimate, except for X1803-245,
where a smaller distance (7.6 kpc) is assumed to keep the maximum
bolometric luminosity below the Eddington limit.

\subsection {PCA Pointed Observations}

For each of the 37 sources listed in Table~\ref{tab:xobs}, we searched
the \xte archive for pointed observations with the PCA instrument
\citep{jah96}. The number of pointings (excluding raster scans) and
the total exposure time are listed in cols. 2 and 3, respectively.  We
analyzed all of the data that were publicly available, 1996 January
through 2004 November for BHs and through 2004 March for NSs. We
constructed light curves at 1 s time resolution for the full bandwidth
of the PCA instrument (effectively 2--40 keV), averaged over the
number of operating PCUs. Standard filters were applied to screen out
data affected by pointing anomalies, Earth occultations, detector
problems, or unusually high rates in the particle background. We
further impose a lower limit of 2 mCrab (5 c s$^{-1}$ PCU$^{-1}$) on
the net flux from the X-ray source to ensure that the source is not in
a quiescent state. The latter step excludes many observations of
GX339-4.

Two types of PCA spectral analyses are performed for this study.  The
first task is to evaluate the source luminosity for each observation,
in the context of searching for bursts as a function of bolometric
luminosity.  For this purpose we use PCA ``standard 2'' data,
where the X-ray spectrum from each PCU is accumulated in 129 energy
channels with 16 s time resolution.  We use the prescribed methods
for background subtraction and the creation of response files,
which are never offset from the time of each observation by more than
20 days. Spectral fits are performed with the XSPEC package provided
by NASA.  The fits are confined to the energy range of 2.9--25.0 keV,
and we impose a 1\% systematic uncertainty on each spectral bin.  We
focus the analyses on PCU \#2 because of its frequent use and 
excellent calibration, as judged from spectral modeling of 
observations of the Crab Nebula.  Results from PCU \#0 are used in 
substitution when needed.

Each data interval (excluding bursts, when present) is fit to a
spectral model consisting of a disk blackbody, a power-law component,
and a Fe~K$\alpha$ emission line. Photoelectric absorption in the
ISM is also included in the model; the column density is fixed for
each source, and the values are given in Table~\ref{tab:xobs}. For the
NS group, the Fe line is modeled with a Gaussian profile, while a
relativistically broadened line (Laor profile) is used for the BH and
BHC groups.  In addition, a reflection component is needed for BHBs
4U~1354-64, 4U~1543-47, and GX339-4.  A smeared absorption edge is
included in the model for GRO~J1655-40 and XTE~J1550-564
\citep{sob00}. To limit systematic problems beyond the range of the
PCA sensitivity, we impose a lower limit of 0.5 keV for the disk
temperature. Finally, for those data intervals where the minimized
value of reduced $\chi^2$ remains above 1.3, we substitute a broken
power-law model in place of the power law, and we accept the revised
fit if there is significant improvement in $\chi^2$, while the break
energy is above 5 keV, so that the break is not convolved with the
thermal component.

The second type of PCA spectral analysis is the effort to model the
spectra of candidate bursts. The events located in burst searches (see
below) are brief, compared to the 16 s quanta of standard 2 data, and
their analysis requires access to the discretionary fast-timing modes
allocated to each \xte observer. Most NS observations include an event
mode configuration with 64 energy channels and a time resolution of
125 $\mu$s or better.  There is greater variety in the user modes for
BHBs and BHCs; the time resolution is almost always better than 8 ms,
but the number of energy channels is more commonly 8 or 16.  For each
burst candidate, we integrate the excess flux above the local burst
``background'', which includes the persistent source emission plus the
various types of celestial and detector backgrounds. The net spectrum
is then modeled with a blackbody function and photoelectric
absorption to determine whether the event is consistent with thermal
emission that is expected from a thermonuclear explosion.  The
mechanics of the fit resemble the procedures used for a source's
persistent emission, albeit with fewer PCA energy channels.

The accumulated time of analyzed PCA data for each source is given in
col. 4 of Table ~\ref{tab:xobs}.  By group, the total exposures are
3.8 Ms for BHBs, 2.7 Ms for BHCs, and 3.7 Ms for NSs.  The 4 sources of
unknown type are ignored in our calculations of burst rates, but their
total archival time is only 0.14 Ms, or 1.3\% of the time utilized for
burst searches.  We conclude that the archive of \xte pointed
observations of our 37 non-pulsing Galactic X-ray transients is highly
concentrated into the source classes where measured burst rates can be
effectively compared to physical models.

\section{X-ray Timing and Spectral Analyses}

\subsection {Candidate Bursts from PCA Light Curves}

We conduct computer searches for X-ray bursts using the 1 s light
curves derived from the PCA instrument.  A candidate burst is defined
via variance analysis in the following way.  For each 1 s time bin
under scrutiny, we define two background intervals with relative time
offsets: $-280 < BG1 < -20$, and $180 < BG2 < 280$.  These
``background'' intervals contain flux contributions from the source's
persistent emission, the diffuse X-ray background, and the detector
background. The asymmetric gap between these background intervals is
designed to contain X-ray bursts, which have rise times of a few
seconds and decay times ranging from 10 s to a couple of minutes
\citep{str06}.

For each background region, we compute the mean and the sample
standard deviation: $b_1, \sigma_1, b_2, \sigma_2$, respectively.  At
a particular time $t_0$, the count rate $C(t_0)$ is tested for a
condition of excess X-ray flux:

\begin{equation}
(C(t_0) - b_1) > 5 \sigma_1 \hskip 0.2cm  ;  \hskip 0.2cm (C(t_0) - b_2) > 5 \sigma_2
\label{eq:bcand}
\end{equation}

Sample light curves containing X-ray bursts and the offset background
regions for burst searches are shown in Fig.~\ref{fig:xburst}.  Both
of these bursts arise from Aql~X--1, and they have durations
that are significantly shorter and longer, respectively,
than the average for that source.

The use of two background regions dampens the rate of false candidates
when the source exhibits secular changes in X-ray flux or state
transitions.  If the available data in a background region are less
than 10 s because of data gaps, then $b$ and $\sigma$ are set to zero,
effectively removing one side of the variance test (Eq. 1).  The use
of sample standard deviations, rather than Poisson statistics, is
required because there is considerable source flickering as is evident
in the power density spectra of X-ray binaries at 1 Hz \citep{vdk06,
mcc06}.

There are two noteworthy details to report regarding the effort to
uniformly search for burst candidates using PCA light curves and
Eq. 1.  First, the NS source X1658-298 exhibits X-ray eclipses and
absorption dips \citep{wij02} that complicate the variance analyses.
This requires hands-on attention to the light curves, and we added one
burst candidate that was not recognized by the automatic search
process.  Secondly, the NS system EXO~1745-248 sometimes exhibits
extraordinary light curves in the form of structured dips that mingle
with X-ray bursts \citep{mark00,wij05}. During such episodes,
the variances in the background regions can become very large,
given the mean count rate.  However, upon review of the results, we
choose to accept the automatic search results for this source.

\begin{figure}
\figurenum{2} \plotone{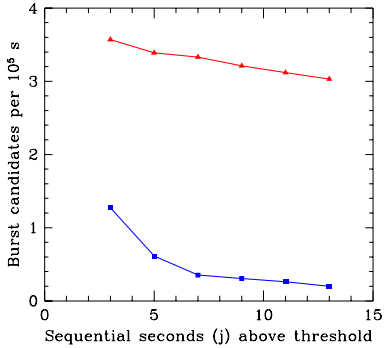}
\caption{Rate of finding burst candidates as a function of the
number of sequential seconds for which the variance test for excess flux
(Eq. 1) is satisfied. Triangles denote the NS group, while squares
represent the combined results for BHBs and BHCs.
\label{fig:jseq}}
\end{figure}

A burst candidate is defined with one additional free parameter, which
is how many consecutive seconds, $j$, that Eq. 1 is true. To
investigate this parameter, we have iterated the search for burst
candidates, varying $j$ over the range 3--13 s.  The results are shown
in Fig.~\ref{fig:jseq}. The rates for finding burst candidates are
shown separately for the NS group (triangles; excluding the eclipsing
source X1658-298) and for the combined BHB and BHC groups (squares).
For the NS group, the rate is high and the distribution is flat versus
$j$. Lower rates are seen for BHs, and the rate decreases more
steeply with $j$, as high-amplitude examples of fast flickering
transition to events with longer time scales that might be X-ray
bursts.

Burst searches should be as complete as possible in order to fairly
compare the burst rates with physical models for thermonuclear
explosions. For the NS group, the number of burst candidates is
insensitive to the choice of $j$ (Fig.~\ref{fig:jseq}).  We therefore
adopt $j = 7$ as a choice that effectively avoids many fast events for
BHBs and BHCs, while preserving the need to implement search criteria
that are as short as the timescale for decay to half-maximum in true
bursts \citep{str06}. This decision is not expected to create a bias
against detection true bursts in BH systems, since our theoretical
models (\S 4) show that burst durations are longer for heavy compact
objects as compared to NS bursts. The number of burst candidates for
each source with $j = 7$ is included in Table~\ref{tab:xobs}.

Before moving on to the method to distinguish true X-ray bursts from
candidates, we examine whether the selection of candidates appears to
be systematically affected by the search parameters.  Some of the
burst-search criteria are scientifically motivated, e.g. the window
function for variance analysis. Other choices (e.g. the $5 \sigma$
threshold) are dictated by statistics and affected by non-burst source
behavior.  It is therefore instructive to compare the detectability of
the burst candidates versus the detection threshold.  As the search
window steps over a given burst, the critical value in determining the
condition of excess X-ray flux is the {\it j}th brightest point,
evaluated in units of signal-to-noise relative to the background
region that imposes the highest threshold.  We can use this quantity
as a detectability parameter: $D_j$ = maximum\{$x_j(t_i)$\}, where
$t_i$ is incremented over the entire burst candidate, $x_j(t) = $
minimum value of \{$r(t), r(t+1), ... r(t+j-1)$\}, $r(t+n) = (C(t+n) -
b_k) / \sigma_k$, and the index $k$ is the one (of two) background
region that minimizes $r(t+n)$.

\begin{figure}
\figurenum{3} \plotone{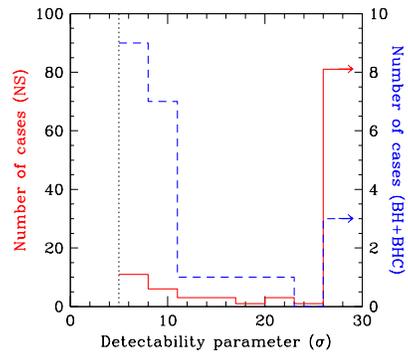}
\caption{ The detectability of burst candidates is quantified by
sweeping the search window over the burst to find the maximum
signal-to-noise ratio of the {\it j}th brightest point above the
background (see text).  These histograms are made for the burst
candidates that were identified using $j = 7$ s.  Most of the burst
candidates for the group of combined BHBs and BHCs (blue, dashed line)
have detectability that is close to the $5 \sigma$ threshold, which is
shown as a dotted vertical line. In contrast, the burst candidates for
neutron stars (red, solid line) have much higher significance. The
horizontal axis is truncated to allow comparison of the two histograms,
and the arrows for the last data points indicate that the plotted 
number includes all cases from the lower edge of the bin (26 $\sigma$) 
to the maximum observed value.
\label{fig:bdetect}}
\end{figure}

The detectability results for burst candidates ($j=7$ s) are shown in
Fig.~\ref{fig:bdetect}. For the NS group we use all of the burst
candidates tabulated in Table~\ref{tab:xobs} except for those from
X1658-298.  The large majority of NS burst candidates have
detectability that is well above the $5 \sigma$ threshold; 86\% have
detectability above $10 \sigma$ and 72\% are above $30 \sigma$.  On
the other hand, the detectability distribution for the BHB and BHC
groups lies closer to the detection threshold, with only 39\% above
$10 \sigma$.  Therefore, at the burst candidate level, we already see
a large difference between the rates and statistical properties of
results for the NS group as compared to BHBs and BHCs.

\subsection{Burst Confirmation from PCA Spectral Analyses}

\begin{figure}
\figurenum{1} \plotone{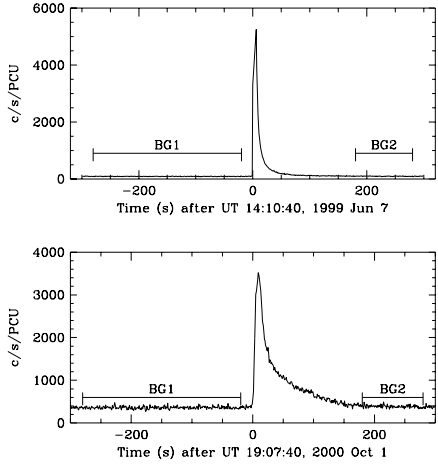}
\caption{Sample of a short (top panel) and long (bottom) X-ray burst
from Aql~X--1. The time references are the burst trigger time, or
the first second where Eq.~1 is true for 7 consecutive s. These
events would trigger bursts for times $> 20$ consecutive s.
\label{fig:xburst}}
\end{figure}

The standard method to identify a burst candidate as a bona fide type
I X-ray burst (i.e. a thermonuclear explosion) is to demonstrate that
the X-ray flux above the persistent level has a thermal (blackbody)
spectrum \citep{str06} with a temperature below 3 keV. The large
majority of PCA observations of the sources in the NS group provide
data telemetry in an event mode with 64 or more spectral channels and
122 $\mu$s or better time resolution, and this facilitates the effort
to isolate the burst candidate and conduct meaningful spectral
analyses.  For all of the burst candidates that we detect
(Table~\ref{tab:xobs}, col. 9) we integrate the excess flux and fit
the net spectrum to a blackbody model, as described in \S 2.2. All but
two of the candidate bursts from the NS group (i.e., 135 of 137) are
confirmed as type I X-ray bursts. We further note that all of these
bursts exhibit smooth temporal profiles that include a fast rise and
slower decay, as illustrated in Fig.~\ref{fig:xburst}, except for some
bursts from EXO~1745-248 and X1658-298 that are co-mingled with
structured absorption dips.

The 20 burst candidates for BHBs and BHCs are significantly fewer than
the NS candidates, while their temporal profiles are substantially more
complicated.  We divide them into two groups in the effort to assess
their viability as thermonuclear explosions from the surface of a
compact object. None of the BH and BHC burst candidates has profiles
that resemble the type I bursts shown in Fig.~\ref{fig:xburst}.
Nevertheless, we assess them primarily on the basis of 
spectral properties, as done for the NS burst candidates.

\begin{figure}
\figurenum{4} \plotone{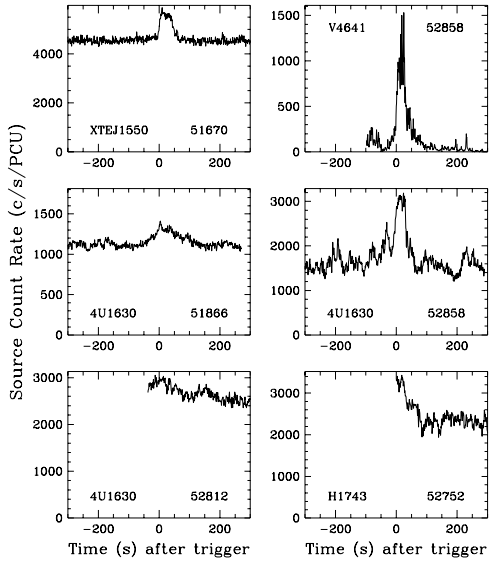}
\caption{Samples of the burst candidates with durations $> 30$ s
for the BHB and BHC groups.  Each panel is labeled by the source name
(abbreviated) and the observation day (MJD).  The time axis is
referenced to the burst trigger time.  None of these events
was confirmed as a type I X-ray burst. In each case the spectrum of 
the excess flux is inconsistent with a blackbody spectrum.
\label{fig:rogues1}}
\end{figure}

In 15 cases, the flare timescale (FWHM) is 30-100 s, and this provides
access to at least one 16-s spectrum from PCA standard 2 mode (129
energy bins) during the peak of the event.  Some illustrations of
these BH burst candidates are given in Fig.~\ref{fig:rogues1}.  For
the majority (9) of long-duration burst candidates, the trigger occurs
near the beginning or the end of the observation, when there is only
one background region available in the burst search process. Two such
examples are shown in the lowest two panels of Fig.~\ref{fig:rogues1}.
Given the chronic drifts in the X-ray flux of BH transients, it seems
obvious that such events are spurious candidates, selected when a
strong local maximum happens to occur near the beginning or end of an
observation. The remaining seven burst candidates with long durations
are perhaps the most interesting, since most of them occur during
bright X-ray states, when heavy compact objects are expected to burst
most frequently (see \S 5). Four illustrative examples are shown in
the upper panels of Fig.~\ref{fig:rogues1}.

For each of the 15 long-duration burst candidates, we subtracted the
spectrum of the persistent emission from the burst maximum, as sampled
in the 16 s quanta of the standard 2 data mode. We then fit the
results to spectral models, as we had done for NS burst
candidates. None of these burst candidate spectra is consistent with a
blackbody spectrum. In most cases the spectral model requires both a
thermal and a power-law component, suggesting a simple flaring of the
persistent emission.  Some of the burst candidates, e.g. those from
XTE~J1550-564, are well fit as simple power-law spectra with photon
indices in the range 2.8-3.7.

\begin{figure}
\figurenum{5} \plotone{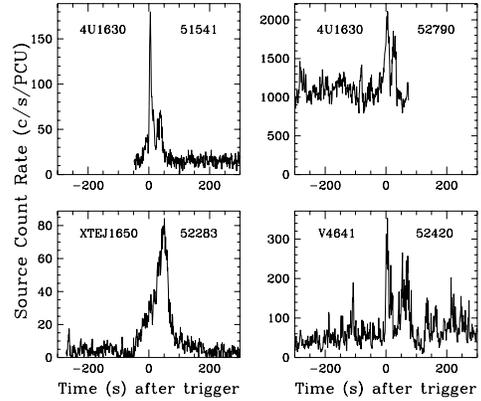}
\caption{Samples of the BH burst candidates with short duration.
The plot labels follow the convention of Fig.~\ref{fig:rogues1}. 
None of these events is a type I X-ray burst that would
signify a thermonuclear explosion on the surface of a compact object.
\label{fig:rogues2}}
\end{figure}

The remaining 5 BH and BHC burst candidates are fast events, and four
examples are displayed in Fig.~\ref{fig:rogues2}. Three candidates
occur in faint X-ray states (persistent count rates $< 100$ c s$^{-1}$
PCU$^{-1}$). The two fast events in bright states are both from
4U~1630-47, and one of these is displayed in the top-right panel of
Fig.~\ref{fig:rogues2}. The spectral analyses of the fast events must
depend on the discretionary telemetry modes that have much higher time
resolution than the standard 2 mode. In these particular cases, the
selected modes provide 24 to 40 PCA energy channels over the range of
2 to 40 keV. For the short duration candidates, we again find that the
burst spectra are inconsistent with a blackbody function and a
temperature below 3 keV.  The results for the faint burst from
XTE~J1650-500 (see Fig.~\ref{fig:rogues2}) are marginal.  However,
Tomsick et al. (2003) \nocite{tom03} conducted a detailed study of
flares from this source during the faint decay of its outburst, and
they concluded that those events are not type I X-ray bursts.

Finally, we note that one can construct a source hardness ratio (after
subtracting the background) from using the PCA count rate at 15-40 keV
relative to the rate at 2-15 keV. Since thermal spectra (1-3 keV)
contribute negligible flux above 15 keV, the plot of this hardness
ratio shows a dip during type I X-ray bursts, while other types of
flares in our study appear either neutral or show increases at times
near the burst trigger.

We conclude that we are unable to find any type I X-ray bursts during
6.5 Ms of PCA exposures of BHBs and BHCs.  It remains to be seen
(\S5.2) whether the exposure times, binned versus Eddington
luminosity, are sufficient to violate the predictions of the burst
model for massive compact objects with a hard surface.

\subsection{X-ray Spectral Differences between Neutron Star and Black Hole Groups}

X-ray spectral fits (\S 2.2) are necessary to compute the
bolometric luminosity for each observation (excluding any bursts),
since this quantity serves as a proxy for the local accretion rate
that is used to derive the burst rate function (\S 5.1).
This large set of spectral parameters also allows us to compare
different source classifications, with a desire to gain corroborative 
evidence for intrinsic differences that would help to justify
the group assignments gained from the literature.

We examined the distributions in the value of the power-law index for
the NS and BH groups, and we find that they are similar.  For each
group, a steep power-law component is seen during soft X-ray states,
with photon index, $\Gamma > 2.5$. The less common hard state shows a
flatter power-law component with $\Gamma \sim 1.7$. Such results are
typical for BHBs \citep{mcc06}, and also for the hard and soft
branches of atoll-type NS systems \citep{mun03,mai04}.

However, the temperature distributions for the thermal component
appear to effectively distinguish the BH and NS groups.  These
measurements are confined to the soft states, since the thermal
component is generally not visible in PCA data (i.e., above 2 keV) in
the hard states of either BHBs \citep{mcc06} or NSs (e.g.,
Gierli{\'n}ski \& Done 2002) . \nocite{gie02} This is shown in
Fig.~\ref{fig:Tdiff}.  The $\sim 1$ keV temperatures for BHs (blue
line) and BHCs (green line) are clearly disjoint from the $\sim 2$ keV
temperature of NS systems.  We note that the NS histogram in
Fig.~\ref{fig:Tdiff} excludes the results from two dipping sources
(X1658-298 and EXO~1745-248), where the temperature distributions may
be affected by variations in the absorption column density.  We also
confined the results for BHs and BHCs to observations in the thermal
state \citep{mcc06}, ignoring the observations in the ``steep power
law'', a spectral state that is not observed for NS systems.

The large majority of X-ray transients do exhibit soft X-ray states.
Exceptions in our sample are limited to two BHBs (XTEJ~1118+480 and
4U~1354-64), where the outbursts are confined to the hard state.  We
conclude that the group assignments of Table 1 can be reproduced with
the combined use of dynamical mass measurements and the empirical
conclusion that heavy compact objects in the thermal state exhibit
mean temperatures below 1.5 keV. Alternatively, sources with mean
soft-state temperatures above 1.5 keV are all associated with the NS
group (including the two dippers). These results provide evidence 
for intrinsic differences between the BH and NS groups that are
independent of the phenomenon of X-ray bursts.  The design of our
statistical study is thereby strengthened.

\section{The X-ray Burst Model}

We describe the burst model that is used to calculate the rates of
type I X-ray bursts for both accreting NSs and heavy compact
objects. In \S5, we compare the model predictions to the measured NS
burst rate function and the upper limits derived for BHBs and BHCs.
We use the theoretical model of Cooper \& Narayan (2005),
\nocite{coop05} which is an expanded and improved version of the model
of Narayan \& Heyl (2003). \nocite{nar03} In the subsections below, we
summarize the assumptions of the model \citep{coop05}, and we describe
the modifications to the model that are relevant to the present study.
  
\subsection{Modifications to the Burst Model for Neutron Stars}
 
We assume that gas accretes spherically onto a NS of
gravitational mass $1.4 M_{\odot}$ and areal radius $10.4$ km.  The
accreted gas is composed of $70\%$ hydrogen, $28\%$ helium, $1.6\%$ CNO
elements, and $0.4\%$ iron by mass.  We assume that the crust either
forms an ordered lattice at high densities or is amorphous.
Furthermore, we assume that the stellar core emits neutrinos via
either modified Urca reactions or pionic reactions.
 
The only difference between the superburst code described by Cooper \&
Narayan (2005)\nocite{coop05} and the model used here to calculate
hydrogen- and helium-triggered bursts is the way in which we calculate
the inner temperature boundary condition.  The long recurrence times
of superbursts ($\sim$ years to decades) imply that the thermal
diffusion depth is very deep into the NS, often deeper than
the crust-core interface.  Therefore, to solve for the thermal and
hydrostatic profiles of the stellar crust, Cooper \& Narayan (2005)
integrated down to the crust-core interface, at which they applied an
inner temperature boundary condition.  However, hydrogen- and
helium-triggered burst recurrence times are much shorter, so the
thermal diffusion depth is much closer to the stellar surface.
Narayan \& Heyl (2003) integrated only to this diffusion depth, at
which they applied their inner temperature boundary condition.  Since
they were unable to model the crust of the star below neutron drip,
they assumed that the star was isothermal below the diffusion depth.
The newer code \citep{coop05} accurately models the entire crust,
rendering this assumption unnecessary.

In an X-ray transient, the time-averaged accretion rate prior to the
transient outburst, $\langle \dot{M} \rangle$, is much lower than the
accretion rate during the burst, $\dot{M}_{\mathrm{burst}}$.  The
energy generation rates due to thermonuclear burning, compressional
heating, and deep crustal heating, which help set the thermal profile
of the crust, are all proportional to the accretion rate.  Since the
diffusion time down to the crust-core interface in a NS is
$\sim 10$-$100$ years, the thermal profile of most of the crust will
be set by the time-averaged accretion rate $\langle \dot{M} \rangle$,
not the current accretion rate $\dot{M}_{\mathrm{burst}}$ during the
burst.  We allow for this as follows in the calculations.
 
To calculate the inner temperature boundary condition for NS
transients, applied at the diffusion depth $\Sigma_{\mathrm{diff}}$,
we first solve for the time-averaged thermal profile of the NS
star crust.  Using the time-averaged accretion rate $\langle \dot{M}
\rangle$, we integrate down to the crust-core interface and apply the
procedure described by Cooper \& Narayan (2005) to calculate the inner
temperature boundary condition at this interface.  The resulting
equilibrium configuration gives a temperature profile as a function of
depth, $T(\Sigma)$.  We then use the burst accretion rate
$\dot{M}_{\mathrm{burst}}$ and proceed with the burst calculation as
described by Narayan \& Heyl (2003), setting the inner temperature
boundary condition to the value of the temperature profile $T(\Sigma)$
evaluated at the diffusion depth $\Sigma_{\mathrm{diff}}$.
 
\subsection{Modifications to the Burst Model for Black Holes}
 
In our black hole burst model, we assume that BHs are compact objects
with solid surfaces.  We assume spherical accretion and the same gas
composition as that used in the NS model.  We assume a gravitational
mass of $M = 8.0 M_{\odot}$, and we choose three stellar areal radii,
$R = 9/8 R_{\mathrm{S}}$, $2 R_{\mathrm{S}}$, and $3 R_{\mathrm{S}}$,
where $R_{\mathrm{S}} = 2GM/c^{2}$ is the Schwarzschild radius.  Note
that $R = 9/8 R_{\mathrm{S}}$ is the physical lower limit on the
radius of a compact object with a solid surface and a nonpositive
density gradient (Buchdahl 1959)\nocite{buc59}.  The maximum stellar
radius is not well constrained, but X-ray observations of BHBs imply
that the inner radius of the accretion disk is $\lesssim 3
R_{\mathrm{S}}$ \citep{dav05,li05,sha06}.  Furthermore, we assume that
the outermost layer of the compact object consists of normal baryonic
matter.  Below this layer we make no assumptions regarding either the
stellar structure or composition. We define the transition rest mass
density $\rho_{0,\mathrm{trans}}$ to be the density that separates
these two regimes.  Thus for $\rho_{0} < \rho_{0,\mathrm{trans}}$ the
structure and composition are that of a NS and for $\rho_{0} >
\rho_{0,\mathrm{trans}}$ the structure and composition are unknown.
 
The BH burst calculation is identical to the NS burst calculation,
with three exceptions: (i) To solve for the time-averaged thermal
profile, we integrate into the compact object until the rest mass
density $\rho_{0} = \rho_{0,\mathrm{trans}}$, whereas in the NS case we
integrate to the crust-core interface, where $\rho_{0} =
2\times10^{14}$ $\mathrm{g\,cm^{-3}}$.  (ii) We do not know up to what
density $\rho_{0,\mathrm{trans}}$ a hypothetical massive compact
object with a surface should consist of normal baryonic matter.  Thus,
we choose three values: $4 \times 10^{10}$$ \mathrm{g\,cm^{-3}}$, $4
\times 10^{11}$ $\mathrm{g\,cm^{-3}}$, and $4 \times 10^{12}$
$\mathrm{g\,cm^{-3}}$.  We find that the results of our calculations
are quite insensitive to the specific value of
$\rho_{0,\mathrm{trans}}$.  Therefore, we choose
$\rho_{0,\mathrm{trans}} = 4 \times 10^{11}$ $\mathrm{g\,cm^{-3}}$,
the neutron drip point, as our fiducial transition rest mass density.
At the depth where $\rho_{0} = \rho_{0,\mathrm{trans}}$, we assign a
fixed inner temperature boundary condition $T_{\mathrm{IBC}}$ instead
of assuming that the interior cools via either modified Urca reactions
or pionic reactions.  Since we make no assumptions about the interior
of the compact object, we cannot determine the proper temperature at
neutron drip self-consistently.  Thus, we choose seven values for
$T_{\mathrm{IBC}}$, covering the range $10^{5}$ K to $10^{8}$ K in
logarithmic steps of 0.5.  These values should generously bracket the
likely true proper temperature.  (iii) The thermal diffusion time to
the depth at which $\rho_{0} = \rho_{0,\mathrm{trans}}$ is
model-dependent, but it is typically on the order of a month.  This is
shorter than the timescale over which the accretion rate changes in a
BH transient, which is on the order of several months.
Therefore, in our model we treat BH transients as
quasi-persistent systems and set $\langle \dot{M} \rangle =
\dot{M}_{\mathrm{burst}}$.  This is different from the NS case.
Nonetheless, we have done BH burst rate calculations in which $\langle
\dot{M} \rangle \ll \dot{M}_{\mathrm{burst}}$, and the results are
generally very similar to those presented in this paper.
 
\section{X-ray Burst Rates and Comparisons with Burst Models}

\subsection{Burst Rates and Model Results for Neutron Stars}

The \xte satellite is in a low-Earth orbit, where astronomical
instruments routinely experience data gaps due to Earth occultations
and electronics shutdown during spacecraft passage through the South
Atlantic Anomaly.  It is therefore highly impractical to directly
measure the time between consecutive X-ray bursts from a given source.
We therefore adopt a different strategy and statistically consider
every continuous data interval as an opportunity to capture X-ray
bursts. We record the exposure time and any burst statistics, and we
extract and fit the average PCA spectrum (excluding any bursts) with
the disk-blackbody + power-law model, as described in \S 2.2.

The exposures and confirmed bursts for all of the X-ray sources in
each compact object group are then accumulated in logarithmic
intervals of the ratio of the bolometric luminosity to the Eddington
luminosity.  For each data interval, we determine the bolometric
luminosity ($L_{\rm bol}$) from the spectral parameters, summing the
contributions from the power law, integrated over 1--30 keV, and the
accretion disk: $L_{disk} = 1.29 \times 10^{35} N_{dbb} (d/10$ kpc$)^2
(T/$keV$)^4 /$ cos($i$) erg s$^{-1}$, where $N_{dbb}$ and $T$ are the
'diskbb' parameters in XSPEC, and the distance ($d$) and binary
inclination ($i$) are listed for each source in
Table~\ref{tab:xobs}. For the Eddington luminosity, we use $L_{\rm
Edd} = 1.3 \times 10^{38} (M/M_{\odot})$ erg s$^{-1}$, where the
compact object mass ($M$) is also given in Table~\ref{tab:xobs}.  We
note that 'diskbb' is a non-relativistic model that is only used to
integrate the thermal spectrum from the accretion disk.

\begin{figure}
\figurenum{7} \plotone{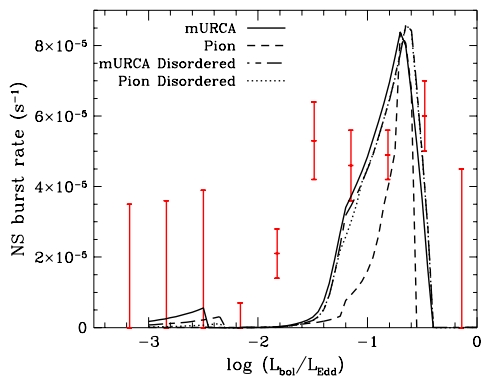}
\caption{Average burst rates (red points; $1 \sigma$ error bars) for
the NS group, binned as a function of the bolometric luminosity
divided by the Eddington luminosity.  The burst models all assume a
mass of 1.4 \msun, radius of $10.4$ km, $\langle \dot{M} \rangle$
equal to $10^{-3} \dot{M}_{\mathrm{Edd}}$, and the distances given in
Table~\ref{tab:xobs}.  The burst rates predicted by four burst models
are shown.  The Pion model is ruled out, but the other three models
are generally consistent with the data.
\label{fig:nsrates}}
\end{figure}

\begin{figure}
\figurenum{8} \plotone{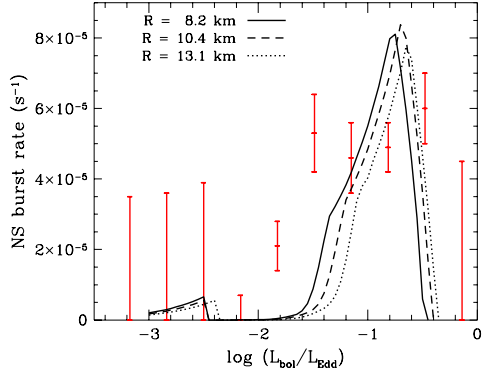}
\caption{Average burst rates for the NS group, as defined for Fig.~6.
Here, the models vary the NS radius, while assuming
a mass of 1.4 \msun, $\langle \dot{M} \rangle$ equal to $10^{-3}
\dot{M}_{\mathrm{Edd}}$, mUrca cooling in the core, an ordered crust,
and the distances given in Table~\ref{tab:xobs}.  The models are all roughly
consistent with the data.
\label{fig:nsratesradii}}
\end{figure}

We refer to the burst rates, measured in logarithmic intervals of
$L_{\rm bol} / L_{\rm Edd}$, as the burst rate function. The results
for the NS group are shown in Figs.~\ref{fig:nsrates} and
\ref{fig:nsratesradii}. The error bars ($1 \sigma$) are calculated by
assuming Poisson statistics for counting bursts in each bin, and the
90\% confidence interval is shown when no bursts were found in a given
bin.  Fig.~\ref{fig:nsrates} shows the predicted burst rates for four
different models of the NS group, all with a gravitational mass of
$1.4 M_{\odot}$ and an areal radius of $10.4$ km.  As explained in \S
4.1, these predictions are derived from models tailored to transient
X-ray sources, where the inner temperature boundary condition for the
NS is set by the long-term average accretion rate of $\langle \dot{M}
\rangle = 10^{-3} \dot{M}_{\mathrm{Edd}}$.  Of the four models shown,
the ``Pion'' model with an ordered crust is inconsistent with the
data.  This model predicts bursts over a very narrow range of
$\dot{M}$ ($\sim$ factor of two), whereas the observations of
individual sources indicate a substantial burst rate over at least an
order of magnitude range in $\dot{M}$.  We may thus rule out this
model for NSs, in agreement with previous studies
\citep{nar03,coop04,brown04,coop05}.  The other three models all give
nearly the same predictions and are generally consistent with the
data.  The predicted burst rate of $\sim 5 \times 10^{-5}$ s$^{-1}$
agrees very well with the observations, as does the position of the
peak as a function of accretion luminosity.  This is significant
because the model has essentially no free parameters.  The range of
luminosity over which the burst rate is substantial is $\sim 0.8$ in
$\log(L_{\rm bol}/L_{\rm Edd})$, which is somewhat lower than the
observed range, which is $\sim 1.2$.  Part of the discrepancy may be
due to uncertainties in the observational estimates of $L_{\rm bol}$
(see below).  Fig.~\ref{fig:nsratesradii} shows the effect of the NS
radius on the predictions.  We see that, over the range $R =
8.2$-$13.1$ km, the models are generally indistinguishable and
consistent with the data.  Narayan \& Heyl (2003)\nocite{nar03} used
their type I X-ray burst models to argue that small radii are
preferred.  The present calculations, which treat the inner boundary
condition more carefully, lead to a different conclusion.

We can roughly characterize the success of the NS burst model in an
integral fashion, as follows. We calculate the total number of
predicted bursts by considering each NS exposure and multiplying the
exposure time by the burst rate given by the model (e.g. mUrca) for
the luminosity on that occasion.  The predicted total is 102 bursts,
while the measured number is 135. We consider this level of 
agreement to be remarkably  good, given the uncertainties in the
true values of the mass accretion rate.

The measured burst rates and the model predictions have similar
values, but the model has a sharper profile, peaking near $L_{\rm bol}
/ L_{\rm Edd} \sim 0.2$.  To understand the significance of this
deviation, we must address the factors of uncertainty in our
determinations of the Eddington-scaled luminosity. For statistical
uncertainties, we roughly estimate that the most significant terms in
the error analysis are a 20\% uncertainty in $M$ and a 30\%
uncertainty in $d$, and a 15\% uncertainty in calculating the
luminosity from the spectral parameters.  This leads to a convolved
statistical uncertainty of 65\% in $L_{\rm bol} / L_{\rm Edd}$ (noting
the quadratic dependence on $d$). There are also important systematic
uncertainties, especially the ability to estimate the mass accretion
rate from the bolometric luminosity.  In view of these uncertainties,
it is not at all surprising that the measured burst function is
broader than the model.

Can the distance uncertainty alone account for the broad shape of the
measured burst function?  We conducted an exercise to find a set of
optimal distances that might bring the measurements into better
agreement with the burst model, without altering any other steps in
the data analyses.  One can view the burst rate measurements, the
statistical uncertainties, and the burst model results (e.g. mUrca) in
Fig.~\ref{fig:nsrates} in terms of a $\chi_{\nu}^2$ test with no
degrees of freedom.  With no considerations of uncertainties along the
luminosity axis, we find $\chi_{\nu}^2 = 5.0$.  The set of optimized
distances for the 13 NSs achieves $\chi_{\nu}^2 = 0.9$, with an average
shift in distance of 3\% and an rms shift of 32\%. If the broad shape
of the burst rate function is, in fact, primarily an effect of
distance errors, then the true distances for Aql X-1 and 4U~1608-52,
which account for the majority of the NS exposure time, must both be
$\sim$47\% larger than the estimates given in Table 1. Such deviations
remain within the realm of possibility, since the primary method for
distance determinations, which utilizes radius-expansion bursts, has
its own imperfections and systematic uncertainties \citep{gal03}.
However, we emphasize that this consideration of the distance
uncertainty is merely an attempt to explore that topic in a
quantitative manner.  The results do not imply 
that other sources of errors should be neglected.

\subsection{Burst Upper Limits and Model Results for Black Holes}

\begin{figure}
\figurenum{9} \plotone{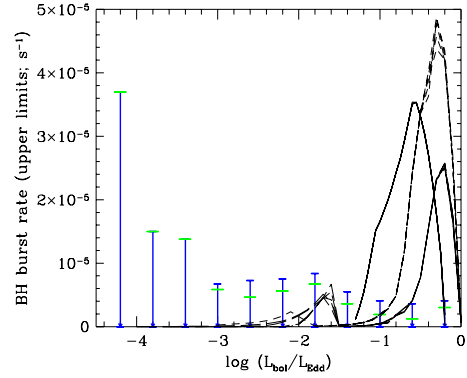}
\caption{Upper limits (90\% confidence) for burst rates in BHBs
(upper/blue error bars), and for the combined BHB and BHC groups
(lower/green error bars), binned as a function of the bolometric
luminosity divided by the Eddington luminosity. The burst rates
predicted for a heavy compact object of mass $8 M_{\odot}$ with a
solid surface are shown by the curves.  The three groups of curves
correspond to different assumptions for the radius: $R/R_{\mathrm{S}}
= 3$ (the shortest peak on the right), $2$ (the tallest middle peak),
and $9/8$ (the intermediate height peak on the left).  In each group
seven models were considered: $T_{\mathrm{IBC}} = 10^{5}$ to $10^{8}$
K, with steps of 0.5 in logarithmic units.  The narrowness of each
group of seven overplotted curves at large values of $L_{\rm
bol}/L_{\rm Edd}$ illustrates the insensitivity of the burst rates to
the inner temperature boundary condition.
\label{fig:bhrates}}
\end{figure}

The measured burst functions for BHB groups are shown in
Fig.~\ref{fig:bhrates}.  For each luminosity bin, there are two upper
limits that correspond to the 90\% confidence values (assuming Poisson
statistics) for the BHB group and the combined BHB and BHC group,
respectively.  Fig.~\ref{fig:bhrates} also shows the predictions of
the burst model for heavy compact objects.  Like the NS model results,
the predicted burst rates for large values of $L_{\rm bol} / L_{\rm
Edd}$, where the rates are maximized, are quite insensitive to the
inner temperature boundary condition.  These bursts ignite very near
the surface of the compact object, which is thermally insulated from
the stellar interior \citep{coop06}, and thus the burst rates are
quite insensitive to the inner temperature boundary condition.  Bursts
at lower accretion rates generally are triggered much deeper in the
ocean, where the thermal profile of the ignition region is more highly
correlated with that of the inner crust.  Consequently, the burst
rates are much more sensitive to the inner temperature boundary
condition in this regime.

As illustrated in Fig.~\ref{fig:bhrates}, the results of all 21 BH
burst models (i.e. 3 stellar radii $\times$ 7 inner temperatures) are
grossly inconsistent with the observed upper limits for $L_{\rm
bol}/L_{\rm Edd} \gtrsim 0.1$.  To quantify the degree to which the
model results are inconsistent with the data, we proceed as follows.
For a particular BH model $\mu$ and assuming Poisson statistics, the
probability of observing zero bursts in an accretion rate bin given
that the model is an accurate description of BHs found in nature is
simply

\begin{equation}
P_{i}(0|\mu) = \exp(-r_{i} t_{i}),
\end{equation}
where $i$ denotes the accretion rate bin, $r_{i}$ is the burst rate
(bursts s$^{-1}$) for bin $i$ calculated using the theoretical model
$\mu$, and $t_{i}$ is the total exposure time for bin $i$.  The
probability of observing zero bursts in all bins given that model
$\mu$ is correct is therefore
\begin{equation}
	P(0|\mu) = \prod_{i} P_{i}(0|\mu) = \exp \left(- \sum_{i}
	r_{i} t_{i} \right).
\end{equation}

We calculate the probability of observing zero bursts given that BHs
have solid surfaces for each of the stellar areal radius and inner
temperature boundary condition models discussed in \S4.  The
theoretical burst rates are quite sensitive to the interior physics of
the compact object for $\log(L_{\rm bol} / L_{\rm Edd}) \leq -1.5$.
Therefore, when we calculate the probability of observing zero bursts,
we consider only those accretion rate bins $i$ such that $\log(L_{\rm
bol} / L_{\rm Edd}) \geq -1.5$.  The results are listed in Table 2.
The probability that our observational measurements are consistent
with the assertion that BHs are compact objects with solid surfaces is
less than $2 \times 10^{-7}$.  When we include BHCs, we obtain an even
more significant result, viz., the probability is less than $3 \times
10^{-13}$ (see Table 3).  We are unable to construct any model of a
compact object with a solid surface composed of normal baryonic matter
that is compatible with observations.

\section{Discussion and Conclusions} 

The absence of type I X-ray bursts in BHBs and BHCs fully supports the
current paradigm: black holes do not have a solid surface that can
accumulate accreting gas and produce the recurrent nuclear explosions
that characterize accretion onto NSs.  This study confirms
the argument proposed by Narayan \& Heyl (2002) and the subsequent
analysis of T03 that the absence of type I X-ray bursts suggests the
presence of a BH event horizon. In the present paper, we
significantly improve the quantification of the discrepancy between
the upper limits for BH bursts and the model predictions for heavy
compact objects with a hypothesized hard surface.  This is done by
choosing sources that spend appreciable time in states of high
luminosity, by explicitly measuring the burst rate as a function of
the estimated accretion rate, and by making improvements to the burst
model.  The luminosity-binned exposure times of the {\em RXTE} sample
of X-ray transients are sufficient to categorically exclude all models
of heavy compact objects with radii and inner temperatures within the
possible ranges for heavy compact objects found in nature, as long as
the accretion conditions conform to the general assumptions of the
burst model.  By combining the luminosity-binned exposure times with
the results of our burst model, we calculate the probability that BHBs
contain heavy compact objects with solid surfaces to be less than $2
\times 10^{-7}$.  For the combined BHB and BHC group, the probability
drops to less than $3 \times 10^{-13}$.

\begin{figure}
\figurenum{6} \plotone{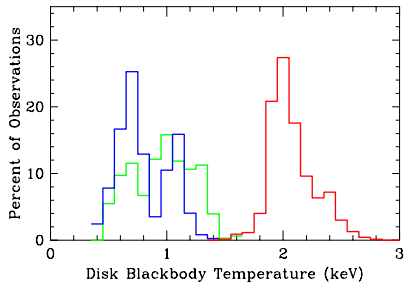}
\caption{Normalized distributions in the temperatures of the thermal
component for different groups of compact objects. Higher temperatures
are seen for the NS group (red line), selecting the soft-state
observations of 11 sources (excluding the two dippers).  The groups of
BHBs (blue line) and BHCs (green line) show systematically lower
temperatures. These data are also selected for soft X-ray color, and they
additionally satisfy conditions of the thermal state given by
McClintock \& Remillard (2006).  This excludes soft-state observations
that are in the ``steep power law'' state, which is a spectral state not
observed in NS systems.
\label{fig:Tdiff}}
\end{figure}

Our compilation of spectral fit parameters allows us to compare NS and
BH groups in terms of properties that have nothing to do with X-ray
bursts.  Systematic differences are found in the temperatures of the
thermal component that is observed during soft X-ray states
(Fig.~\ref{fig:Tdiff}). The hotter characteristic temperatures of NS
systems is likely to originate in the boundary layer that forms when
accreting matter reaches the NS surface. This is another consequence
of the NS hard surface that is being investigated for implications of
BH event horizons \citep{don03}, as noted in the \S1.  In the context
of our X-ray burst study, we confirm the kinship between dynamical
BHBs and BHC, as well as the separate identity of all the known
bursters.

The burst model performs well in predicting the number of type I
bursts in our sample of NS transients. The results agree
reasonably well with the burst rates $\sim 7.7 \times 10^{-5}$ s$^{-1}$
measured by Cornelisse et al. (2003) \nocite{cor03} with the {\it
Beppo}-SAX Wide Field Camera for three active sources.  Our measured
burst rate function is broader than the model predictions, but
significant broadening is expected due to the uncertainties in the
distance to each binary system.  Systematic uncertainties may further
contribute to any discrepancies. The burst rate depends, in part, on
the mass accretion rate, which we calculate from a simple scaling of
the bolometric luminosity.  While this process appears to be
reasonable for X-ray states dominated by a thermal spectrum, it is
uncertain whether the non-thermal states operate with similar
radiative efficiency.  This includes the hard (steady jet) and ``steep
power-law'' states of BH systems \citep{mcc06}, and an analogous hard
state for NS transients, as well.  However, we note that the
exposure times for each type of compact object are dominated by the
soft X-ray states where thermal radiation contributes most of the
emitted energy.  This is especially true at high luminosity, where the
BH groups display gross inconsistencies between the model burst rates
and the observed upper limits for thermonuclear explosions.  Thus, the
uncertainties in the mass accretion rate do not seriously weaken our
conclusion that BHs lack a solid surface.

Finally, one must question whether our modeling efforts for the
hypothesized massive compact objects cover all of the options
available to nature. Our results seriously restrict the possibility
that black holes have a surface composed of normal baryonic matter.
However, the results do not constrain the viability of more exotic
BH models (e.g. Chapline et al. 2001; Robertson \& Leiter
2002; 2003; Abramowicz, Kluzniak, \& Lasota 2002, Yuan, Narayan, \&
Rees 2004).  \nocite{cha01,rob02,rob03,abr02,yua04} There are similar
concerns for our assumption of a spherical accretion geometry, and the
implicit presumption that any accretion-focusing effects
(e.g. magnetic field) would cause observable pulsations (for objects
with surfaces) that are not observed for BHBs and BHCs.  As noted in
\S1, the best strategy remains to search for solid surface effects in
BHBs with as many techniques as possible.  Nevertheless, we have shown
in this paper that type I X-ray bursts provide strong indirect
evidence for BH event horizons, and it would appear that the evidence
can be refuted only by invoking rather exotic physics.
 
\acknowledgements Partial support was provided by the NASA contract to MIT 
for RXTE instruments, and NASA grant NNG04GL38G to R.~N.

\clearpage  

\begin{deluxetable}{lrrrrrcccc}
\tablecaption{Non-pulsing Galactic X-ray Transients: 1996--2004}
\tablecolumns{10}
\tablehead{
\colhead{X-ray} & \colhead{No.} & \colhead{archive} & 
\colhead{anal.} & \colhead{$d$} & \colhead{mass} & 
\colhead{$N_H$} & \colhead{$i$} & \colhead{cand.} & 
\colhead{type I} \\
\colhead{name} & \colhead{obs.} & \colhead{(ks)\tablenotemark{a}} & 
\colhead{(ks)\tablenotemark{b}} & 
\colhead{(kpc)} & \colhead{(\msun)} & \colhead{($10^{22}$)} & 
\colhead{(deg.)} & \colhead{bursts\tablenotemark{c}}  & \colhead{bursts} \\
} 
\startdata 
\cutinhead{Black Holes (BHBs)}
Black Holes (BHBs) & & & & & & & & \\

XTE~J1118+480 &  53  &  152 &  143  &  1.8 &  6.8 & 0.012 &  70 & 0 & 0 \\
4U~1354--64   &  10  &   67 &   65  & 27.0 & 10.0 &  1.1  &  60 & 0 & 0 \\
4U~1543--47   & 109  &  271 &  260  &  7.5 &  9.4 &  0.4  &  21 & 0 & 0 \\
XTE~J1550-564 & 421  & 1233 & 1030  &  5.3 &  9.6 &  0.9  &  72 & 3 & 0 \\
XTE~J1650-500 & 183  &  358 &  345  &  7.0 &  5.5 &  0.67 &  60 & 2 & 0 \\
GRO~J1655-40  &  81  &  388 &  377  &  3.2 &  6.3 &  0.9  &  70 & 0 & 0 \\
GX339-4       & 586  & 1749 & 1061  &  4.0 &  7.0 &  0.6  &  50 & 0 & 0 \\
V4641~Sgr     &  68  &  172 &  154  &  9.8 &  7.1 &  var. &  66 & 2 & 0 \\
XTE~J1859+226 & 133  &  386 &  365  & 11.0 &  9.8 &  0.4  &  60 & 0 & 0 \\
\cutinhead{BH Candidates (BHCs)}
BH Candidates (BHCs) & & & & & & & & & \\
4U~1630-47       &  595 & 1294 & 1156 & 8.5 & 8.0 &  8.8  &  70 & 8 & 0 \\
H~1743-322       &  256 &  906 &  822 & 8.5 & 8.0 &  2.2  &  60 & 5 & 0 \\
SAX~J1711.6-3808 &   18 &   47 &   45 & 8.5 & 8.0 &  2.8  &  60 & 0 & 0 \\
XTE~J1720-318    &  102 &  301 &  288 & 8.5 & 8.0 &  1.2  &  60 & 0 & 0 \\
GRS~1739-278     &   11 &   30 &   27 & 8.5 & 8.0 &  1.7  &  60 & 0 & 0 \\
SLX~1746-331     &   51 &  149 &  145 & 8.5 & 8.0 &  1.8  &  60 & 0 & 0 \\
XTE~J1748-288    &   24 &  106 &  103 & 8.5 & 8.0 &  8.0  &  60 & 0 & 0 \\
XTE~J1755-324    &    4 &   12 &   10 & 8.5 & 8.0 &  0.6  &  60 & 0 & 0 \\
XTE~J1908+094\tablenotemark{d}    &   32 &   41 &  0.0 & \nodata & \nodata& \nodata & \nodata & \nodata & \nodata \\
XTE~J2012+381    &   24 &   60 &   58 & 8.5 & 8.0 &  0.7  &  60 & 0 & 0 \\
\cutinhead{Neutron Star Bursters}
Neutron Star Bursters & & & & & & & & & \\
4U~1608-52       & 378 & 1451 & 1387 &  3.6 & 1.4 &  1.0 &  60 & 26 & 26 \\
X1658-298        &  79 &  355 &  351 & 10.0 & 1.4 &  2.0 &  77 & 26\tablenotemark{e} & 26 \\
RX~J1709.5-2639  &  38 &  142 &  139 &  8.5 & 1.4 &  0.3 &  60 &  3 &  3 \\
X1711-339        &  10 &   42 &   39 &  7.5 & 1.4 &  1.5 &  60 &  0 &  0 \\
XTE~J1723-376    &   2 &   20 &   19 &  8.5 & 1.4 &  4.0 &  72 &  2 &  2 \\
SAX~J1747.0-2853 &   9 &   42 &   41 &  7.5 & 1.4 &  8.8 &  60 &  0 &  0 \\
EXO~1745-248     &  52 &  149 &  145 &  8.7 & 1.4 &  3.0 &  60 & 24 & 22 \\
X1744-361        &  10 &   39 &   38 &  8.5 & 1.4 &  2.5 &  60 &  0 &  0 \\
X1745-203        &  18 &  130 &  128 &  8.5 & 1.4 &  0.6 &  60 & 16 & 16 \\
SAX~J1750.8-2900 &  12 &   48 &   47 &  5.2 & 1.4 &  2.0 &  60 &  4 &  4 \\
X1803-245        &  11 &   35 &   33 &  7.6 & 1.4 &  2.0 &  60 &  0 &  0 \\
SAX~J1810.8-2609 &   0 &    0 &    0 &  \nodata & \nodata &  \nodata & \nodata & \nodata & \nodata \\
AQL~X-1          & 297 & 1297 & 1258 &  5.0 & 1.4 &  0.3 &  60 & 31 & 31 \\
XTE~J2123-058    &   5 &   67 &   66 &  9.6 & 1.4 & 0.06 &  60 &  5 &  5 \\
\cutinhead{Unclassified}
Unclassified & & & & & & & & & \\
SAX~J1428.6-5422 &  18 &   37 &    0 &  \nodata & \nodata &  \nodata & \nodata & \nodata & \nodata \\
XTE~J1739-285    &   7 &   21 &    0 &  \nodata & \nodata &  \nodata & \nodata & \nodata & \nodata \\
SAX~J1805.5-2031 &  45 &   70 &    0 &  \nodata & \nodata &  \nodata & \nodata & \nodata & \nodata \\
XTE~J1856+053    &   2 &    8 &    0 &  \nodata & \nodata &  \nodata & \nodata & \nodata & \nodata \\ 
\enddata  
\tablenotetext{a}{\xte total archive, as of 2004 November 1.}
\tablenotetext{b}{We analyzed the portion of the archive that is publicly available, shows a source flux above 5.0 c s$^{-1}$ PCU$^{-1}$, and is free of systematic problems.}  
\tablenotetext{c}{These triggers are based on timing analyses, with a 
   requirement for 7 sequential seconds that a burst candidate is above 
   the $5 \sigma$ trigger threshold (see text).}
\tablenotetext{d}{Source excluded since there is another known X-ray
   source in the PCA field of view.}  
\tablenotetext{e}{One burst was added because X-ray eclipses and absorption dips 
   interfere with the variance test for detecting burst candidates.}
\label{tab:xobs}  
\end{deluxetable}     

\newpage

\begin{deluxetable}{c|ccc}
\tablewidth{0pt}
\tablecaption{Probabilities of observing no bursts from the group of
BHBs}
\tablehead{
\colhead{$T_{\mathrm{IBC}}$} & 
\colhead{$R/R_{\mathrm{S}} = 9/8$} &
\colhead{$R/R_{\mathrm{S}} = 2$} &
\colhead{$R/R_{\mathrm{S}} = 3$}
}
\startdata
$10^{5.0}$ K&$3\times10^{-16}$&$1\times10^{-16}$&$1\times10^{-7}$\\
$10^{5.5}$ K&$3\times10^{-16}$&$1\times10^{-16}$&$1\times10^{-7}$\\
$10^{6.0}$ K&$3\times10^{-16}$&$1\times10^{-16}$&$1\times10^{-7}$\\
$10^{6.5}$ K&$3\times10^{-16}$&$6\times10^{-17}$&$9\times10^{-8}$\\
$10^{7.0}$ K&$3\times10^{-16}$&$3\times10^{-17}$&$1\times10^{-7}$\\
$10^{7.5}$ K&$4\times10^{-16}$&$5\times10^{-17}$&$1\times10^{-7}$\\
$10^{8.0}$ K&$4\times10^{-16}$&$4\times10^{-17}$&$1\times10^{-7}$\\
\enddata
\end{deluxetable}

\begin{deluxetable}{c|ccc}
\tablewidth{0pt}
\tablecaption{Probabilities of observing no bursts from the combined
group of BHBs and BHCs}
\tablehead{
\colhead{$T_{\mathrm{IBC}}$} & 
\colhead{$R/R_{\mathrm{S}} = 9/8$} &
\colhead{$R/R_{\mathrm{S}} = 2$} &
\colhead{$R/R_{\mathrm{S}} = 3$}
}
\startdata
$10^{5.0}$ K&$3\times10^{-39}$&$5\times10^{-35}$&$2\times10^{-13}$\\
$10^{5.5}$ K&$3\times10^{-39}$&$5\times10^{-35}$&$2\times10^{-13}$\\
$10^{6.0}$ K&$3\times10^{-39}$&$5\times10^{-35}$&$2\times10^{-13}$\\
$10^{6.5}$ K&$3\times10^{-39}$&$2\times10^{-35}$&$1\times10^{-13}$\\
$10^{7.0}$ K&$3\times10^{-39}$&$5\times10^{-36}$&$2\times10^{-13}$\\
$10^{7.5}$ K&$4\times10^{-39}$&$1\times10^{-35}$&$2\times10^{-13}$\\
$10^{8.0}$ K&$4\times10^{-39}$&$1\times10^{-35}$&$2\times10^{-13}$\\
\enddata
\end{deluxetable}

\newpage

\newpage
 
\newpage

\newpage
 
\newpage

\newpage

\newpage

\newpage

\newpage

\end{document}